\documentclass{article}
\usepackage{graphicx}

\newcommand{\bfr}{\begin{flushright}}
\newcommand{\efr}{\end{flushright}}
 
\begin{document}
\title{A Model of BPS Black Holes in a Discrete Space
}
\author{Nahomi Kan\footnote{kan@gifu-nct.ac.jp}
\\
{\small
Gifu National College of Technology,
Motosu-shi, Gifu 501-0495, Japan
}
\bigskip
\\
Koichiro Kobayashi\footnote{m004wa@yamaguchi-u.ac.jp}
\quad
and
\quad
Kiyoshi Shiraishi\footnote{shiraish@yamaguchi-u.ac.jp}\\
{\small
Yamaguchi University,
Yamaguchi-shi, Yamaguchi 753--8512, Japan}
}
\date{\today
}
\maketitle
\begin{abstract}
We examine a model of BPS black holes lying on a discrete extra space.
The geometry is obtained from the discretization of the
harmonic equation. We study the scattering amplitudes of two types of
scalar fields, which correspond to fields in a bulk and on a brane.
We conclude that the two types of scattering can be distinguished in the
region of large transfer momentum.
\end{abstract}

\section{Introduction}
Extension of gravity theory is regarded as an important subject to
study in modern theoretical physics. Some extended models are expected
to be relevant to the alternative of dark contents in the universe
\cite{Clif,Capo}. Also in a microscopic perspective, modification of the
Einstein gravity is motivated in the community of theoretical
physicists; the theory with good quantum behavior and some natural
explanation to the hierarchical scales in particle physics are eagerly
pursued by many authors.

Now a days, the study of the models of gravity in higher
dimensions, with and without higher-derivative terms in the action,
has been a common topic in theoretical high-energy physics.
Moreover, a broad range of possibilities is investigated,
such as, scalar-tensor theory of gravity, vector-tensor theory,
DBI-type, Lorentz-symmetry-breaking, non-local, and so on.

Massive gravity%
\footnote{For reviews, see \cite{Hin,deR}.}
 is an interesting model for the modified gravity,
because a massive graviton is a natural generalization in particle
physics in popular sense but massive gravity turns out to have many
difficulties as a quantum field theory.
It is known that the construction of ghost-free massive
gravity \cite{RGT,HRS} is naturally derived from bigravity, which has
been studied for several decades \cite{fg,big}. Interestingly, the
generalization of bigravity  may permit multigravity \cite{MG}, and it
is closely related with deconstruction of gravity. The
dimensional deconstruction \cite{ACG,HPW} is an idea of making a
higher-dimensional theory from the lower-dimensional copious fields.
Thus, the dimensional deconstruction can be regarded as a modified and
restricted version of the discretization of space, which assumes the
minimal scale in the length scale. The idea of the smallest length in
our universe has been considered  for a long time as a solution of
removing divergences in quantum field theory. So we have seen here, the
various theories of gravity are mutually related.

It is essential to study the consequence of
the generalized gravity theory with discreteness or other modifications
at strong gravity, because it is known that the weak
gravitational field limit is well-described by Einstein's general
relativity. Therefore the solutions of the gravity theories
which represent for gravitating localized objects are important
theoretical arenas to investigate the feature of gravitation. 
Especially, the interaction with matter fields at strong gravity can
be thoroughly studied if the exact solution of the spacetime geometry is
obtained.

In the present work, we will examine a simple model of a BPS black hole
with a discrete space. In general, the object possessing the BPS
relation in its mass and charges is governed by simple equation of
motion, and is usually motivated by string theory and theories with
supersymmetries. The BPS equation considered here is the Laplace
equation, thus, the discretization of the differential equation can be
done rather in a straightforward way. In this paper, we introduce the
graph Laplacian to perform the discretization. Therefore, the extension
to the general structures of discrete spaces associated with generic
graphs will be possible, though only the simplest case is treated in the
present paper.

The plan of the present paper is as follows. In Sec.~2, we first review
BPS black hole solutions%
\footnote{Strictly speaking, the solution obtained in the BPS limit has
the singularity in the Einstein frame, except for the
Reissner-Nordstr\"om solution ($\alpha=0$).} in the
Einstein-Maxwell-dilaton theory. Subsequently, we introduce the graph
Laplacian to discretize the BPS equation and show its solution in the
simplest case. In Sec.~3, we study the scattering amplitudes of scalar
fields with the BPS black hole with a discretized extra space. We treat
them in the Born approximation in the present paper. We consider two
types of scalar fields, one is obtained from the discretization of the
continuum theory, another is the field living in one site of the
discrete space. We concentrate ourselves on finding the way
how we can `see' the black hole by different kinds of scalar fields.
Section~5 is devoted to summary and outlook.

\section{BPS black holes and discretization}
In this section, we review the simplest system allowing a BPS solution,
the Einstein-Maxwell-dilaton theory.
The action for the model in $D$-dimensional (continuous)
spacetime is given by \cite{GM,GHS,KS}
\begin{equation}
S=\int d^Dx\frac{\sqrt{-g}}{16\pi}\left[R-\frac{4}{D-2}(\nabla
\phi)^2-e^{-4\alpha\phi/(D-2)}F^2\right]\,,
\label{a1}
\end{equation}
where $R$ denotes the scalar curvature, $\phi$ stands for the dilaton
field, and the field strength is defined as
$
F_{\mu\nu}=\partial_\mu A_\nu-\partial_\nu A_\mu
$
with an abelian gauge field $A_\mu$.

In this action, $\alpha$ is the dilaton coupling. The
effective massless field theory of string theory can be obtained if one
set as $\alpha=1$. Then, the appropriate scaling of the metric yields
the following `stringy' action:
\begin{equation}
\tilde{S}=\int
d^Dx\frac{\sqrt{-\tilde{g}}}{16\pi}e^{-2\phi}\left[\tilde{R}+4(\tilde{
\nabla}
\phi)^2-\tilde{F}^2\right]\,,
\end{equation}
with $\tilde{g}_{\mu\nu}=e^{\frac{4}{D-2}\phi}g_{\mu\nu}$.
Now, we turn to use the original metric (\ref{a1}) in the following
discussion.

The static BPS solution can be derived, with the following ans\"atze:
\begin{eqnarray}
& &ds^2=g_{\mu\nu}dx^\mu
dx^\nu=-V^{-2(D-3)/(D-3+\alpha^2)}dt^2+V^{2/(D-3+\alpha^2)}d{\bf x}^2\,,
\\
& &e^{-4\alpha\phi/(D-2)}=V^{2\alpha^2/(D-3+\alpha^2)}\,,
\\
& &A_\mu dx^\mu=\sqrt{\frac{D-2}{2(D-3+\alpha^2)}}(1-V^{-1}) dt\,,
\end{eqnarray}
as the solution of the Laplace equation for $V(x^i)$:
\begin{equation}
\partial^2 V=0\,,
\label{lap}
\end{equation}
where the Laplacian is
\begin{equation}
\partial^2\equiv\partial_i\partial^i\,.\qquad (i=1,\dots,D-1)
\end{equation}

Here we `discretize' the equation (\ref{lap}) by replacing a part of
the Laplacian with a certain graph Laplacian. 
We adopt
\begin{equation}
\partial^2\longrightarrow
\sum_{i=1}^d\partial_i\partial^i-a^{-2}\Delta(G)\,,
\end{equation}
where $a$ is a scale of length.
The graph Laplacian $\Delta(G)$ has been introduced in spectral graph
theory \cite{BH,GR,CRS,JMP}. 

\begin{figure}[ht]
\begin{center}
\includegraphics[height=3cm]{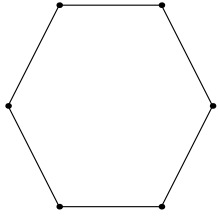}
\end{center}
\caption{A cycle graph, $C_6$.
\label{fig1}
}
\end{figure}

A graph consists of vertices (or sites) and edges
which link two vertices.
For example, a cycle graph $C_N$ has $N$ vertices and $N$ edges
connecting vertices  circularly (Fig.~\ref{fig1}).
A matrix is defined according to the manner of connections of edges to
vertices, and is called a graph Laplacian. For example, the graph
Laplacian of $C_6$ is written as
\begin{equation}
\Delta(C_6)=\left(
\begin{array}{rrrrrr}
2 & -1 & 0& 0&0& -1\\
-1 & 2 &-1 &0 &0&0\\
0 & -1 & 2 & -1 & 0&0\\
0 & 0  & -1 & 2 & -1&0\\
0 & 0  &0& -1 & 2 & -1\\
-1 & 0&0 & 0 & -1 & 2
\end{array}
\right)\,.
\end{equation}

The equation characterizing the eigensystem of the matrix is
\begin{equation}
\Delta(G) v^{(\ell)}=\lambda^{(\ell)}v^{(\ell)}\,,
\end{equation}
where $\lambda^{(\ell)}$ is an eigenvalue of $\Delta(G)$
and $v^{(\ell)}$ is an eigenvector belonging to the eigenvalue.

For $G=C_N$, one can find that
the eigenvalues
\begin{equation}
\lambda^{(\ell)}=4\sin^2\frac{\pi\ell}{N}
\end{equation}
and the eigenvectors
\begin{eqnarray}
v^{(\ell)}&=&(v^{(\ell)}_0, v^{(\ell)}_1,\dots, v^{(\ell)}_k,\dots,
 v^{(\ell)}_{N-1}
)^T\nonumber \\ &=&\frac{1}{\sqrt{N}}(1, e^{i\frac{2\pi\ell}{N}},\dots, 
e^{i\frac{2\pi\ell}{N}k},\dots, e^{i\frac{2\pi\ell}{N}(N-1)})^T\,,
\end{eqnarray}
where $\ell=0, 1, \dots, N-1$. Note that the normalization of inner
products can be fixed as
\begin{equation}
v^{(\ell)}\cdot v^{(\ell')}=\delta_{\ell\ell'}\,.
\end{equation}

Now we turn to the (partial) discretization of the Laplace equation.
We restrict ourselves on the case with the cycle graph $C_N$, hereafter.
Note that the similar discussion for general graphs is possible.
The function $V$ should be interpreted as the functions associated with
vertices of a graph. By using the eigenvectors, we can expand 
\begin{equation}
V_k=\sum_{\ell=0}^{N-1}V^{(\ell)}v^{(\ell)}_k\,,
\end{equation}
where $N$ is the number of eigenvectors, $k$ stands for the
$k$-th vertex and
$V^{(\ell)}$ is the function of $x^i$, $i=1,\dots, d$. Assuming a
`localized' source, at the zero-th vertex, the equation should read
\begin{equation}
\left[\sum_{i=1}^d\partial_i\partial^i-a^{-2}\Delta(C_N)\right]V_k=-4\pi\mu
{N}\delta^d(x^i)\delta_{k0}\,.
\end{equation}

We find the solution for $d=3$:
\begin{equation}
V_k(r)=1+\frac{\mu}{r}\sum_{\ell=0}^{N-1}\exp\left[-2\left|\sin
\frac{\pi\ell}{N}\right|\frac{r}{a}\right]e^{i\frac{2\pi\ell}{N}k}\,,
\label{sol}
\end{equation}
which seems to be a sum of the Newtonian potential and the Yukawa-type
potentials.

Taking the limit of a large number of vertices,
such that $N\rightarrow\infty$ and a small discretized scale
$a\rightarrow 0$ while
$L\equiv Na$ is constant, with introducing a continuous parameter
$y\equiv ka$ ($0\le y<L$), we obtain
\begin{eqnarray}
& &\sum_{\ell=0}^{N-1}\exp\left[-2\left|\sin
\frac{\pi\ell}{N}\right|\frac{r}{a}\right]e^{i\frac{2\pi\ell}{N}k}
\rightarrow
\sum_{\ell=-\infty}^{\infty}\exp\left[-
\frac{2\pi|\ell|}{N}\frac{r}{a}\right]e^{i\frac{2\pi\ell}{N}k}\nonumber
\\
& &=\frac{\sinh(2\pi r/L)}{\cosh(2\pi r/L)-\cos(2\pi y/L)}\,.
\end{eqnarray}
This expression has appeared when we considered the BPS black holes
in the Kaluza-Klein compactification on
$S^1$ with the circumference $L$
\cite{SorK}.

Incidentally, the solution (\ref{sol}) can be expressed by the infinite
sum as
\begin{eqnarray}
& &\sum_{\ell=0}^{N-1}\exp\left[-2\left|\sin
\frac{\pi\ell}{N}\right|\frac{r}{a}\right]e^{i\frac{2\pi\ell}{N}k}\nonumber
\\
& &=
\sum_{q=-\infty}^{\infty}\left\{\frac{Nr}{\pi
a}\left[(Nq-k)^2-\frac{1}{4}\right]^{-1}
{}_1F_2(1;3/2-(Nq-k),3/2+(Nq-k);r^2/a^2)\right.\nonumber \\
&
&\qquad+\left.\frac{N}{\cos[(Nq-k)\pi])}I_{2(Nq-k)}(2r/a)\right\}\,.
\end{eqnarray}
The every part in the parentheses has the limit
\begin{equation}
\{\cdots\}\stackrel{N\rightarrow\infty,
Na=L}{\longrightarrow}\frac{Lr}{\pi}
\frac{1}{r^2+(Lq-y)^2}\,,
\end{equation}
which is the Green's function in four-dimensional space with mirror
sources. Thus the every part in the parentheses corresponds to the
Green's function in three-dimensional space and on-dimensional infinite
lattice.

\section{Scattering amplitudes of scalar waves}
Scattering by black holes has been studied in the
literature, such as Ref.~\cite{FHM}. In this section, we treat the
scattering only in the simple way, since the model we consider here is
still a toy model. We consider scattering of scalar wave in four
dimensional spacetime, i.e.,
$d=3$.
Suppose that the wave equation is assumed as
\begin{equation}
\sum_{i=1}^3\partial_i\partial_i\psi+p^2\psi-U(r)\psi=0\,,
\end{equation}
where $U(r)$ is an effective potential for scattering and $p$ stands
for the momentum of the incident wave.

It is well known \cite{Messiah} that the Born approximation leads to 
the form of the following scattering amplitude $f(\theta)$,
where $\theta$ denotes the scattering angle, 
\begin{equation}
f(\theta)=-\frac{1}{q}\int_0^\infty r\, U(r) \sin qr \,dr
\end{equation}
with the transfer momentum
\begin{equation}
q=2p\sin\frac{\theta}{2}
\end{equation}
and $p$ is the wave number of the incident wave.
The scattering cross-section is simply given by
\begin{equation}
\frac{d\sigma}{d\Omega}=|f(\theta)|^2\,.
\end{equation}

We will consider two types of scalar fields and how the black hole
described by the solution (\ref{sol}) can be seen by the waves in the
Born approximation.
\subsection{Massless scalar field in the `bulk'}
We first consider the scalar field originally defined in
$D$-dimensional spacetime and define its discretized version, which
corresponds to the scalar plane wave spreading in the bulk space. 
Thus, we shortly call this type of scalar field as the `bulk' scalar.

The wave
function of the massless scalar in continuum
$D$-dimensional spacetime is written as
\begin{equation}
\frac{1}{\sqrt{-g}}\partial_\mu(\sqrt{-g} g^{\mu\nu}\partial_\nu\psi)
=0
\end{equation}
If the background geometry with the solution (\ref{sol}) is substituted
and monochromatic wave 
$\psi\propto e^{-i\omega t}$
is presumed, the wave equation takes the
form
\begin{equation}
\partial^2\psi+V^{2(D-2)/(D-3+\alpha^2)}\omega^2\psi=0\,.
\end{equation}
Note that for the massless field, $\omega=p$.
For we consider a scalar field in the bulk now, we adopt the lowest
eigenstate as the $d+1$-dimensional massless scalar. We assume, that is,
\begin{equation}
\psi\rightarrow\psi^{(0)}\,,\quad \Delta(G)\psi^{(0)}\approx 0\,.
\end{equation}
Therefore the wave equation becomes
\begin{equation}
\sum_{i=1}^d\partial_i\partial_i\psi^{(0)}_k
+V_k^{2(D-2)/(D-3+\alpha^2)}\omega^2\psi^{(0)}_k=0\,.
\end{equation}

To use the Born approximation for $d=3$, we should use the trace of the
matrix element of state vector from $k=0$ to $k=N-1$. 
Thus, the scattering amplitude is given by
\begin{equation}
f(\theta)=-\frac{1}{q}\int_0^\infty r\, U(r) \sin qr \,dr
\end{equation}
with
\begin{equation}
U(r)=-\frac{\omega^2}{N}\sum_{k=0}^{N-1}
\left(V_k^{2(D-2)/(D-3+\alpha^2)}-1\right)\,.
\end{equation}

\begin{figure}[ht]
\begin{center}
\includegraphics[height=3cm]{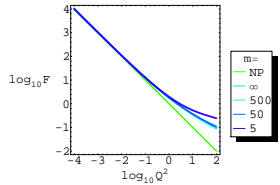}
\end{center}
\caption{The scattering amplitude of the `bulk' scalar for $N=6$.
From top to the bottom on the right-hand side of the curves
correspond to $m=5, 50, 500, \infty$ and the case with pure Newton
potential, respectively.
\label{fig2}
}
\end{figure}

\begin{figure}[ht]
\begin{center}
\includegraphics[height=3cm]{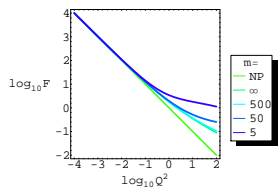}
\end{center}
\caption{The scattering amplitude of the `bulk' scalar for $N=30$.
From top to the bottom on the right-hand side of the curves
correspond to $m=5, 50, 500, \infty$ and the case with pure Newton
potential, respectively.
\label{fig3}
}
\end{figure}

One can find a special case.
If we consider the case with the dilaton coupling $\alpha=1$, then
we find
\begin{equation}
U(r)=-\frac{\omega^2}{N}\sum_{k=0}^{N-1}
\left(V_k^{2}-1\right)\,,
\end{equation}
and this is independent of the dimensionality $D$.
Substituting the solution of $V_k$ for $C_N$ obtained in the previous
section, we get
\begin{equation}
-\frac{1}{\omega^2}U(r)=\frac{2\mu}{r}+\frac{\mu^2}{r^2}\sum_{\ell=0}^{N-1}
\exp\left[-4\left|\sin
\frac{\pi\ell}{N}\right|\frac{r}{a}\right]\,,
\end{equation}
and this leads to the following scattering amplitude:
\begin{equation}
\frac{1}{\omega^2}f(\theta)=\frac{2\mu}{q^2}+\frac{\mu^2}{q}\sum_{\ell=0}^{N-1}
\arctan\frac{aq}{4\left|\sin
\frac{\pi\ell}{N}\right|}\,.
\end{equation}
The amplitude as the function of the transfer momentum in this case is
shown in Fig.~\ref{fig2} for $N=6$ and 
Fig.~\ref{fig3} for $N=30$.

Here we define
normalized quantities $F\equiv\left|\frac{f(q)}{2\mu^3\omega^2}\right|$
and $Q^2\equiv\mu^2q^2$.
In each figures, curves for several different values of $m\equiv \mu/a$
are indicated. The line indicated as `NP' represents for the
amplitude by the pure Newton potential for a reference case. Note the
variables defined here will be used throughout the present paper. 

Since the constant $\mu$ indicates the size of the `black hole',
the next leading contribution to the Newtonian potential can be detected
at large $Q$. When the scale of discreteness (or the minimal length) $a$
is sufficiently small compared with $\mu$, the dependence on $N$ becomes
small.

\begin{figure}[ht]
\begin{center}
\includegraphics[height=3cm]{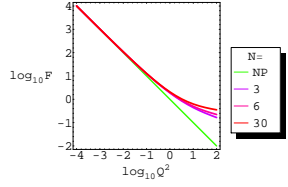}
\end{center}
\caption{The scattering amplitude of the `bulk' scalar for $\mu/(Na)=1$.
From top to the bottom on the right-hand side of the curves
correspond to $N=30, 6, 3$ and the case with pure Newton
potential, respectively.
\label{fig4}
}
\end{figure}

\begin{figure}[ht]
\begin{center}
\includegraphics[height=3cm]{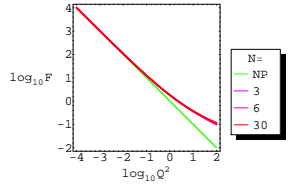}
\end{center}
\caption{The scattering amplitude of the `bulk' scalar for
$\mu/(Na)=10$. From top to the bottom on the right-hand side of the
curves correspond to $N=30, 6, 3$ and the case with pure Newton
potential, respectively.
\label{fig5}
}
\end{figure}

The limit of large $N$ and small $a$ should yields the case with
a compactified space
$S^1$ in the continuum theory, in this case with $C_N$.
The amplitude for a fixed $\mu/(Na)$ is shown in Fig.~\ref{fig4}
for $\mu/(Na)=1$ and Fig.~\ref{fig5}
for $\mu/(Na)=10$.
In each figure, curves for $N=3, 6, 30$ are plotted.

We can find that if the scale of the `extra dimension' $Na$ is small
compared with the black hole scale $\mu$, the discreteness of the
space is difficult to be detected. 

\subsection{Massless scalar field confined on the `brane'}
Next we consider the scalar field living in the single vertex at which
the source of the BPS black hole located.  This is a mimicker of the
field confined on a brane. 
Thus, we may abbreviatedly  call this type of scalar field as the
`brane' scalar. The
corresponding brane is identified with
$k=0$ vertex. The wave equation for the scalar field $\Psi_0$ at the
zeroth vertex is
\begin{equation}
\frac{1}{\sqrt{-q}}\partial_a(\sqrt{-q} q^{ab}\partial_b\Psi_0)
=0\,,
\end{equation}
where $q_{ab}$ ($a, b=0,1,\dots, d$) denotes the $d+1$-dimensional
metric defined through
\begin{equation}
q_{ab}dx^a
dx^b=-V_0^{-2(D-3)/(D-3+\alpha^2)}dt^2+V_0^{2/(D-3+\alpha^2)}\sum_{i=1}^d
d{x^i}^2\,.
\end{equation}
Assuming $\Psi_0\propto e^{-i\omega
t}$ again, we get the following wave equation:
\begin{equation}
\sum_{i=1}^d\partial_i
V_0^{(d+1-D)/(D-3+\alpha^2)}\partial_i\Psi_0+V_0^{(d+D-3)/(D-3+\alpha^2)}
\omega^2\Psi_0=0\,.
\end{equation}
Further we rewrite the equation by using the new variable
\begin{equation}
\Psi_0=V_0^{-\frac{d+1-D}{2(D-3+\alpha^2)}}\psi_0\,,
\end{equation}
as
\begin{eqnarray}
& &\sum_{i=1}^d\partial_i\partial_i\psi_0+
V_0^{2(D-2)/(D-3+\alpha^2)}\omega^2\psi_0\nonumber\\
& &-\frac{d+1-D}{D-3+\alpha^2}\left[
\frac{d-3D+7-2\alpha^2}{2(D-3+\alpha^2)}\sum_{i=1}^d\frac{\partial_i
V_0}{V_0}\frac{\partial_i
V_0}{V_0}+\sum_{i=1}^d\frac{\partial_i\partial_i
V_0}{V_0}\right]\psi_0=0\,.
\end{eqnarray}
Although this equation depends on the the number of spatial dimensions
$d$ as well as the total dimensionality $D$, the last term can be
neglected compared with the second term if we consider sufficiently
high-energy scattering.

For $d=3$ and $\alpha=1$, the wave equation reads at high energy as
\begin{equation}
\sum_{i=1}^3\partial_i\partial_i\psi_0+
V_0^{2}\omega^2\psi_0
\approx 0\,,
\end{equation}
where
\begin{equation}
V_0(r)=1+\frac{\mu}{r}\sum_{\ell=0}^{N-1}\exp\left[-2\left|\sin
\frac{\pi\ell}{N}\right|\frac{r}{a}\right]\,.
\end{equation}
Therefore the effective potential is given by
\begin{equation}
U(r)=-{\omega^2}
\left(V_0^{2}-1\right)\,.
\end{equation}

By the Born approximation, we obtain the following scattering amplitude:
\begin{eqnarray}
& &\frac{1}{\omega^2}f(\theta)=2\mu
\frac{Na^2 \big[{{\big({\sqrt{4+{a^2q^2}}}+aq\big)}^N}+{{\big({\sqrt{4
+{a^2q^2}}}- aq\big)}^N}\big]}{{\sqrt{4 {{a
}^2}{q^2}+{a^4q^4}}} \big[{{\big({\sqrt{4 +{a^2q^2}}}+
aq\big)}^N}-{{\big({\sqrt{4 +{a^2q^2}}}- aq\big)}^N}\big]}
\nonumber \\
&
&\qquad\qquad+\frac{\mu^2}{q}\sum_{\ell_1=0}^{N-1}\sum_{\ell_1=0}^{N-1}\arctan
\frac{aq}{2\left(\left|\sin
\frac{\pi\ell_1}{N}\right|+\left|\sin
\frac{\pi\ell_2}{N}\right|\right)}\,.
\end{eqnarray}

\begin{figure}[ht]
\begin{center}
\includegraphics[height=3cm]{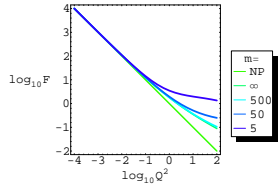}
\end{center}
\caption{The amplitude of the `brane' scalar for $N=6$.
From top to the bottom on the right-hand side of the curves
correspond to $m=5, 50, 500, \infty$ and the case with pure Newton
potential, respectively.
\label{fig6}
}
\end{figure}

\begin{figure}[ht]
\begin{center}
\includegraphics[height=3cm]{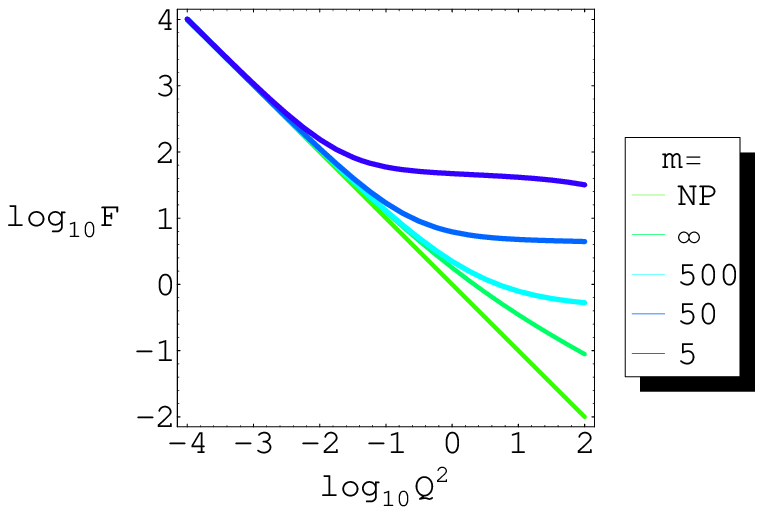}
\end{center}
\caption{The amplitude of the `brane' scalar for $N=30$.
From top to the bottom on the right-hand side of the curves
correspond to $m=5, 50, 500, \infty$ and the case with pure Newton
potential, respectively.
\label{fig7}
}
\end{figure}

The amplitude in this case is shown in Fig.~\ref{fig6} for $N=6$ and in
Fig.~\ref{fig7} for $N=30$.
For the `brane' scalar, the dependence on $m=\mu/a$ is large for large
$N$.

\begin{figure}[ht]
\begin{center}
\includegraphics[height=3cm]{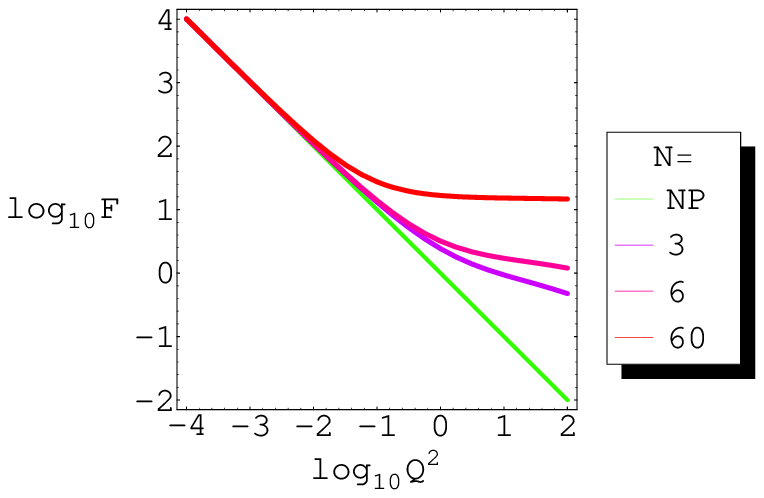}
\end{center}
\caption{The amplitude of the `brane' scalar for $\mu/(Na)=1$.
From top to the bottom on the right-hand side of the curves
correspond to $N=30, 6, 3$ and the case with pure Newton
potential, respectively.
\label{fig8}
}
\end{figure}

\begin{figure}[ht]
\begin{center}
\includegraphics[height=3cm]{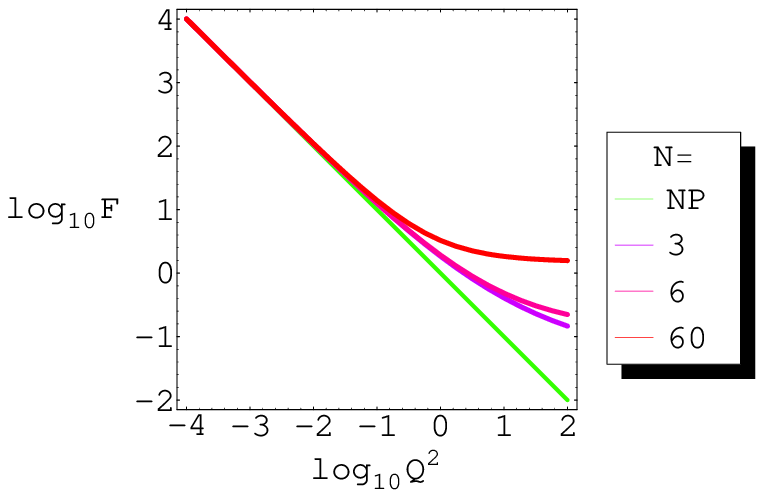}
\end{center}
\caption{The amplitude of the `brane' scalar for $\mu/(Na)=10$.
From top to the bottom on the right-hand side of the curves
correspond to $N=30, 6, 3$ and the case with pure Newton
potential, respectively.
\label{fig9}
}
\end{figure}

The amplitude for fixed $\mu/(Na)$ is shown in Fig.~\ref{fig8} for
$\mu/(Na)=1$ and in Fig.~\ref{fig9} for $\mu/(Na)=10$.
Even if the scale of compactification $\mu/(Na)$ is large,
the dependence on $N$ is not so small.

Since the `bulk' scalar couples only to the Newtonian potential at the
leading order, the dependence on $N$ is rather small. This is because
$V^{(0)}\propto \mu/r$ and the incident wave $\psi\propto \psi^{(0)}$. 

On the other hand, The `brane' scalar couples to every mode, thus the
amplitude is sensitive to all the ratios of variables.

\section{Summary and outlook}
To summarize: We consider the discretization of the BPS equation and
obtain a solution in a simple case, which has a continuum limit of $S^1$
compactification. The solution in the present paper has three length
scales: the radius of the black hole
$\approx
\mu$, the discretization scale
$a$, and the scale of the `extra dimension' $Na$.
The scattering of scalar fields has been studied. The dependence on the
ratio of the variables differs by the type of scalar
fields, the `bulk' scalar and the `brane' scalar.
The `bulk' scalar of the Kaluza-Klein zero mode couples to $1/r$
potential at the lowest order in $\mu$, the dependence of amplitude on
$N$ is rather small. On the other hand,
the `brane' scalar couples to all the components of the potential from
the black hole, therefore the amplitude has large dependence on $N$.

The further study and straightforward extensions of the present work
are expected as follows. The higher-order in the approximation or
numerical derivation of the scattering amplitude should be checked.
The solution describing multi-black holes can also be obtained and
the scattering by the multi-black holes can be calculated.
The use of other graphs than $C_N$ is of importance, such as a path
graph
$P_N$, which imitates $\sim S^1/Z_2$ in the continuum limit.
The graph with vertex weights is analogous to a warped space and is
worth examining. The discretization using disconnected graphs seems to
be a possible non-trivial extension.

We also notify that the scattering by the stringy BPS black hole, in the
case with $\alpha=1$, is independent of the spatial dimensionality.
This is important, if we consider generalization of the model using the
complex graphs. The graph structure has, in general, no continuum
limit in a naive sense as the case considered in the present paper
($C_N\rightarrow S^1$).  As the field theory on fractal graphs have been
studied \cite{Hill1,Hill2},  the gravity on fractal graphs is an
exciting subject to study. The fractal has a unusual dimension, or
in some cases, no uniquely-defined dimension. The stringy case or
special choice of the coupling is substantial in the study of
theory with the fractal (graph). 

The present approach is based on the discretization of the equation of
motion, thus the action of the complete theory has not been considered
yet. In other words, the discretization in our approach is only valid
for the case with a special BPS relation among mass and charges.
Although the investigation into the general case is important, the
BPS case may be  a special point in the `running' couplings and to
study the deviation from the point may be effective at some energy
scale.

Finally, we notify that a discrete object with a certain symmetry is 
interesting from a mathematical point of view, and the model of
magnetic monopole has been considered recently
\cite{Kemp}. Anyway, investigation on the possible sub-structure of our
spacetime should be continued with various approaches.


\end{document}